\def\bold#1{\setbox0=\hbox{$#1$}%
      \kern-.02em\copy0\kern-\wd0
      \kern.04em\copy0\kern-\wd0
      \kern-.02em\raise.0433em\box0 }

\def\bdsmall#1{\setbox0=\hbox{$#1$}%
      \kern-.015em\copy0\kern-\wd0
      \kern.03em\copy0\kern-\wd0
      \kern-.015em\raise.0233em\box0 }

\documentstyle[preprint,amsfonts,prc,aps]{revtex}

\begin{document}
\draft
\title{Nuclear transparency in quasielastic $A(e,e'p)$: \\
intranuclear cascade versus eikonal approximation}
\author{Ye.S. Golubeva}
\address{Institute of Nuclear Research, 117312 Moscow, Russia}
\author{L.A. Kondratyuk}
\address{Institute of Theoretical and Experimental Physics, 117259
Moscow, Russia}
\author{A. Bianconi}
\address{Dipartimento di Chimica e Fisica per i Materiali e per
l'Ingegneria, \\
Universit\`a di Brescia, I-25133 Brescia, Italy, \\
and Istituto Nazionale di Fisica Nucleare, Sezione di Pavia,
I-27100 Pavia, Italy}
\author{S. Boffi and M. Radici}
\address{Dipartimento di Fisica Nucleare e Teorica, Universit\`a di
Pavia, \\
and Istituto Nazionale di Fisica Nucleare, Sezione di Pavia,
I-27100 Pavia, Italy}
\date{\today}
\maketitle
\begin{abstract}
The problem of nucleon propagation through the nuclear medium in
quasielastic $A(e,e'p)$ reactions is discussed in the kinematic range $1
\leq Q^2 \leq 7$ (GeV/$c$)$^2$. Experimental data are available from SLAC,
BATES and, recently, also from TJNAF. The coefficient of nuclear
transparency is calculated for each $Q^2$ in the framework of the
intranuclear cascade model (INC) and of the eikonal approximation (EA).
The former has the capability of directly implementing the detector
acceptances giving a very detailed analysis of the different observables.
The latter, essentially based on an exclusive mechanism, contains
explicit information about the dependence on the target
shell structure. The predictions of both models are in good agreement with 
each other. The INC model reproduces the experimental data quite well in 
the measured range. The EA gives an explanation of the $Q^2$ behaviour of 
the transparency coefficient as a kinematic effect related to the
superposition of contributions from each target shell.
\end{abstract}


\section{Introduction}
\label{sec:intro}

The issue of nucleon propagation through the nuclear medium as a
major problem in understanding nuclear reactions has received much
attention during the last decades. The best tool of investigation is
probably given by an electromagnetic probe knocking out a nucleon from
the nucleus $A$, such as in $A(e,e'p)$ reactions under quasielastic
kinematic conditions~\cite{frumou,bgprep}. In this case, the whole
nuclear volume is explored, the elementary electron-proton scattering
cross section is well known, and high resolution experiments allow for a
clean detection of ejected protons under several kinematic conditions. 

At intermediate energies much work has been done, both theoretically
and experimentally (see, e.g., Ref.~\cite{bgprbook} for a review), and
final-state interactions (FSI) of the ejected proton with the residual
$A-1$ system seem to be well described by an optical potential within
the distorted-wave impulse approximation (DWIA). For large enough
$Q^2 = q^2 - \omega^2$, where $\omega$ and ${\bold q}$ are the energy and
momentum transferred by the electron to the target, respectively,
perturbative QCD predicts the socalled phenomenon of color
transparency~\cite{bro82,mue82,bromue88}, i.e. for increasing $Q^2$ the
struck hadron should propagate undergoing a decreasing interaction with
the nuclear environment. Consequently, the detected proton would emerge
under conditions asymptotically approaching the predictions of the
plane-wave impulse approximation (PWIA) (see
Refs.~\cite{framilstri94,niko94,bbkha94} for a review). 

Experiments have been performed recently at SLAC~\cite{NE18-1,NE18-2}
and TJNAF~\cite{cebaf91013exp}. The SLAC data have been taken in the
range $1 \leq Q^2 \leq 7$ (GeV/$c$)$^2$ and their $Q^2$ and
$A$ dependence do not show conclusive evidence that the transparency
increases with $Q^2$. The new data from TJNAF at $0.64 \leq  Q^2 \leq
3.3$ (GeV/$c$)$^2$ are in reasonable agreement with the prior data from
SLAC. A variety of models have been proposed to describe either the 
evolution of color neutral and compact hadron configurations leading to 
color 
transparency~\cite{bro82,mue82,bromue88,farr88,jenmil91,benhar91,framilstri92,kolya92,kolya94}, 
or the nuclear transparency of proton propagation using conventional
degrees of freedom in the Glauber 
model~\cite{nikoetal94,nikoetal95,nikoetal96}. The data do not rule out 
the possibility of a slow onset of color
transparency, but conventional explanations of nuclear transparency (NT)
have to be first investigated in detail. In fact, this has been done in
Ref.~\cite{NE18-2} within the classical Glauber model and the effective
nucleon-nucleon (NN) total cross section in nuclear medium $\sigma_{\rm
eff}$ has been found lower than the free one $\sigma_{\rm free}$ by $\sim
30\%$. Some reduction of the NN cross section in nuclear medium is indeed
expected from Pauli blocking and short-range
correlations~\cite{vijpie92} as well as from quantum interference
between coherent and incoherent
rescatterings~\cite{nikoetal94,nikoetal95,nikoetal96}. 

The aim of this paper is twofold. We shall first try to study the NT 
occurring during the motion of the ejected proton in terms
of a quasiclassical solution of the multiple scattering. Our approach
will adopt the intranuclear cascade model (INC), a model successfully
developed for the description of hadron-nucleus collisions at
intermediate energies~\cite{gol92,botv} and recently
extended~\cite{gol96,cass,gol97} to account for the in-medium effects in
the production of vector mesons on  nuclei. Then, the results will be
compared with experimental data and with the predictions of the standard 
eikonal approximation (EA)~\cite{glau59}, which has been tested and shown 
to give results for exclusive $A(e,e'p)$ reactions at 
$0.8 \leq Q^2 \leq 4$ (GeV/$c$)$^2$ in remarkable agreement, over a wide 
range of proton angles, with the predictions based on the well 
established optical potential approach~\cite{br95,br96a,br96b}. 

In Sec.\ \ref{sec:geninc} the INC model is applied to quasielastic
semi-inclusive $A(e,e'p)$ reactions. After a brief presentation of the
model in Sec.\ \ref{sec:inc}, momentum and angular PWIA distributions of
the final electrons and protons at different $Q^2$ and for different
targets are generated in Sec.\ \ref{sec:pwia} taking advantage of the
model capability to give a very detailed analysis of different
observables with direct inclusion of detector acceptances, in contrast
to more conventional analytical approaches. The effects of FSI are
discussed in Sec.\ \ref{sec:fsi}. In Sec.\ \ref{sec:eik} a brief review
of EA is presented together with a definition of NT suitable
for comparison with a semi-inclusive measurement. The 
theoretical cross section, essentially based on an exclusive
mechanism, takes into account only the channels related to the
direct proton knockout. However, unlike other semi-inclusive
calculations~\cite{nikoetal94,nikoetal95,nikoetal96}, it
contains explicitly a detailed information on the target shell
structure. Results of the INC model
are compared with data and with the EA prediction in Sec.\
\ref{sec:output}. Some conclusions are presented in Sec.\ \ref{sec:end}.

\section{Quasielastic $A(\lowercase{e,e'p})$ within the 
framework of INC}
\label{sec:geninc}

In the following, a general description of the INC model is
given and angular and energy distributions of generated 
events are discussed for the quasielastic semi-inclusive 
$A(e,e'p)$ reaction on several nuclei. 

\subsection{The INC model}
\label{sec:inc}

The INC model was originally applied to the analysis of hadron-nucleus
interactions~\cite{gol92,botv}. It can be considered as a quasiclassical
numerical representation of the multiple scattering series. It differs
from the standard Glauber approximation~\cite{glau59} in the
description of the multiple incoherent scattering terms. In the latter,
with the socalled frozen approximation, the motion of the scattering 
centers is neglected, while the INC model takes it into account
explicitly. 

Within the INC framework the linearized kinetic equation for the 
many-body distribution function, describing hadron transport in nuclear
matter~\cite{bun}, is solved numerically  by assuming that during the
evolution of the cascade the properties of the target nucleus remain
unchanged. This implies that the number of cascade particles $N_c$ 
is much less than the number of nucleons $A$ in the target nucleus. In
the case of light nuclei this condition might be violated at proton
momenta larger than 5 GeV/$c$, where events with large multiplicities
could be overestimated. This condition does not prevent the application
of the INC model to the description of SLAC~\cite{NE18-1,NE18-2} and 
TJNAF~\cite{cebaf91013exp} data. 

Another feature of the INC model is the fact that the model is
quasiclassical. This might appear a limitation because, consequently, it
cannot describe genuine quantum effects such as the coherent
rescattering. If those effects would be important in the case of
quasielastic $A(e,e'p)$ reactions the application of the INC model would
be doubtful. The struck nucleon, after receiving a large $Q^2$, can in
principle scatter on the residual system in a coherent and incoherent
way. However, in practice in the present kinematic conditions it
cannot transfer a small momentum to the recoiling system because of
Pauli blocking. Therefore, the coherent rescattering is expected to be
suppressed, and incoherent rescattering can adequately be described
by the INC model. On the contrary, within the conventional Glauber
approximation the coherent rescattering is usually overestimated, because
its probability is the same as for the incoherent one and its weight is 
determined by the free NN total cross section.

Within the INC model the target nucleus is regarded as a mixture 
of degenerate neutron and proton Fermi gases in a spherical potential
well with a diffuse surface. The momentum distribution  of the nucleons
is treated in the local density approximation for  a Fermi gas. The
nucleus is divided into a series of concentric zones which help to
follow the propagation of each produced particle from one zone to
another. At the beginning of the cascade a large sample of struck
nucleons is generated. It corresponds to the kinematic conditions of
the quasielastic peak, when the energy $E_e$ of the initial electron 
and $Q^2$ are fixed by the experiment. In this case the scattering angle
for free elastic electron-proton scattering is also fixed. The momentum
and angular distributions of final electrons and struck protons, 
created in any given zone, are determined by the Fermi momentum 
distributions in the same zone. If necessary, final cuts in the momentum
and angular distributions can be applied according to experimental 
acceptances. The relative numbers of struck nucleons produced in
different zones are proportional to the local densities. 

The model describes in a straightforward way the development of the
cascade when the struck proton rescatters elastically or  produces any
number of additional particles, such as pions. Between different
collisions the particles propagate along straight-line trajectories and
the location of the next collision is generated assuming a weight
function exponentially decreasing with the propagation distance. At each
step of the cascade, the competition among different channels is
governed by the channel cross sections, which are taken as in vacuum apart
from the effect of the Pauli exclusion principle taken into account at 
each collision. This means that rescattering may occur only when 
the momentum of each recoiling nucleon is out of the Fermi sphere. In
other words, the damping of the ejectile flux along its trajectory is
determined not by the free NN total cross section, but by a smaller
effective cross section due to Pauli blocking. In principle, the
number of protons counted in the detector includes also those created
with different momenta and subsequently modified by rescattering and
eventually appeared within the momentum acceptance of the detector. These
events, which correspond to many-fold elastic rescattering, are also
taken into account by the INC mechanism.

Masses, energies and momentum components for all the particles in the
initial, final and any intermediate step of the cascade, are recorded
for every event and any necessary distribution involving those quantities
can be produced.   

\subsection{Momentum and angular distributions in PWIA}
\label{sec:pwia}

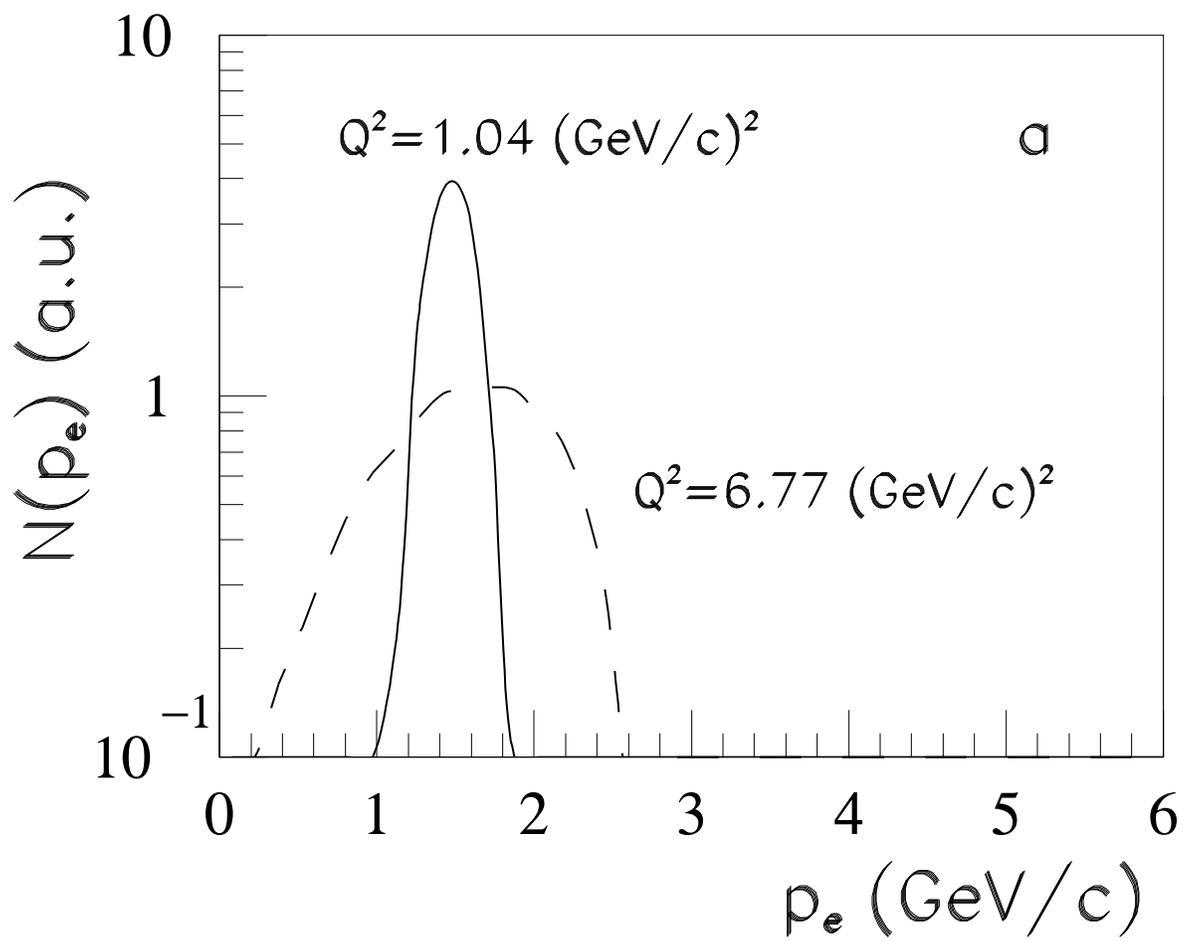
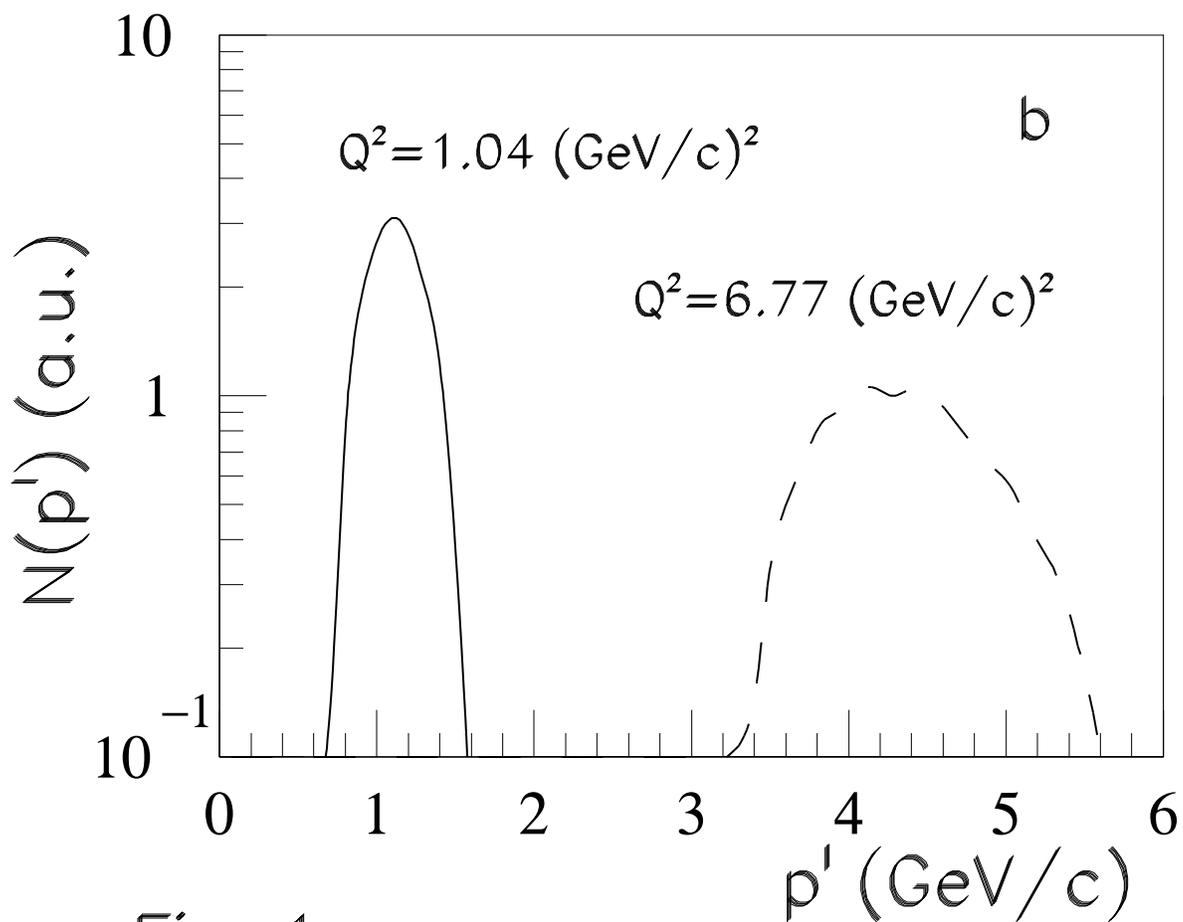
\begin{figure}
\caption{
Momentum distributions of events generated by the INC model in
PWIA for final electrons (a) and protons (b) in the $^{12}$C$(e,e'p)$
reaction in the kinematics of the NE18
experiment.
Solid lines refer to  $Q^2=1.04$
(GeV/$c$)$^2$, $E_e=2.015$ GeV, and dashed lines to $Q^2=6.77$ 
(GeV/$c$)$^2$, $E_e=5.12$ GeV, respectively. }
\label{fig1}
\end{figure}

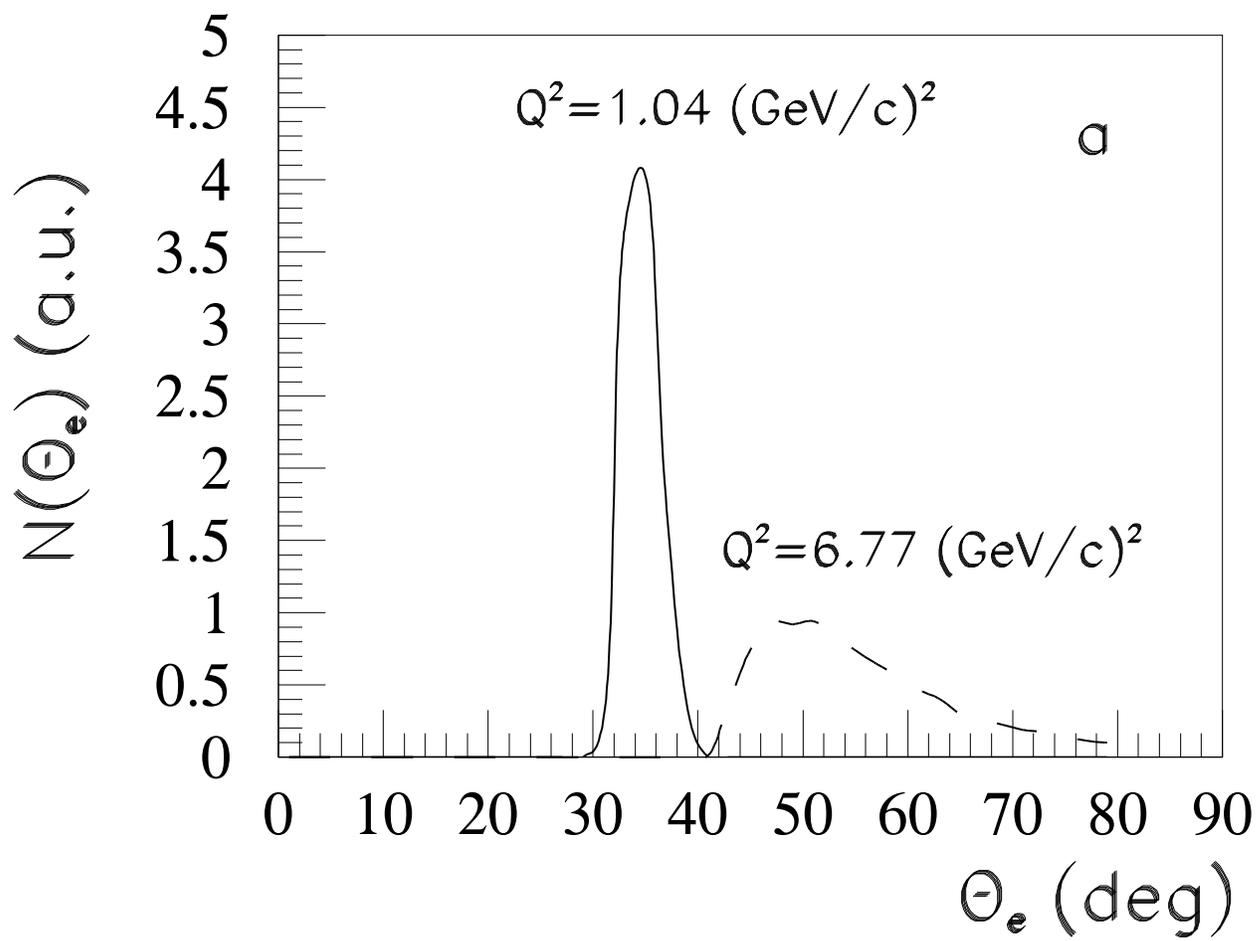
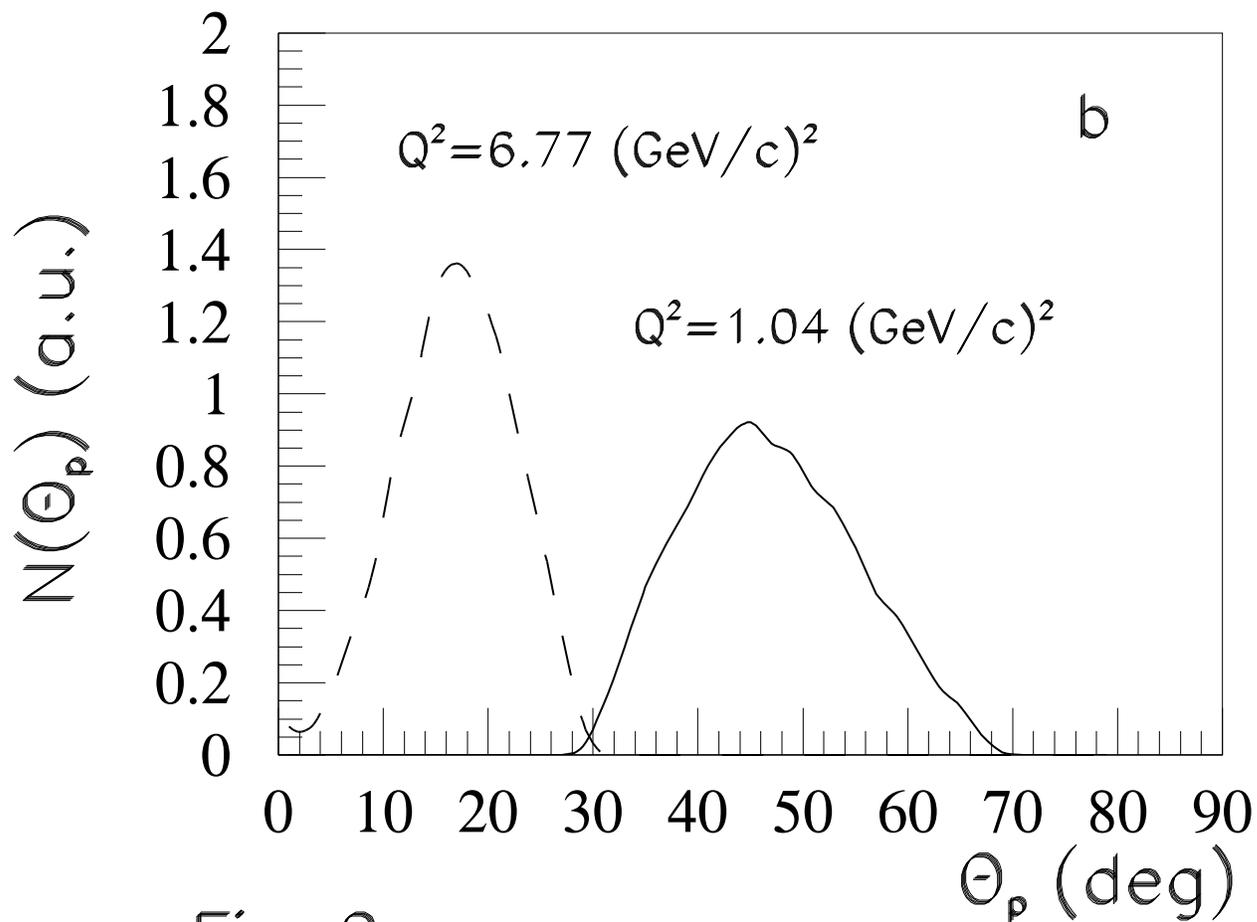
\begin{figure}
\caption{Angular distributions of events generated by the INC model
in PWIA for final electrons (a) and protons (b) in the electron
scattering plane and in the same conditions and notations as in Fig.\
\protect{\ref{fig1}}.}
\label{fig2}
\end{figure}

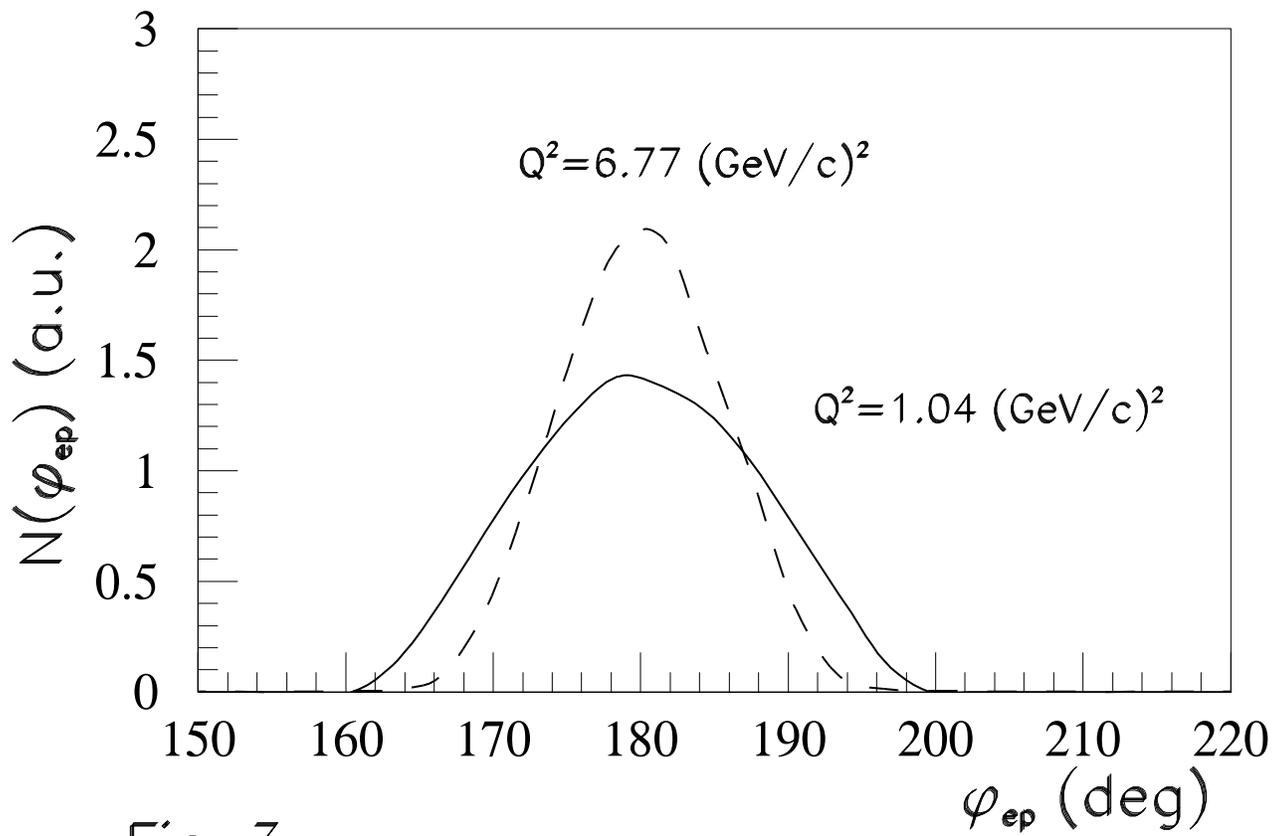
\begin{figure}
\caption{Angular distributions of out-of-plane events generated by the
INC model in PWIA as a function of the angle $\phi_{ep}$ between the
electron scattering plane and the plane defined by the proton and beam
momenta. Kinematics and notations as in Fig.\ \protect{\ref{fig1}}.}
\label{fig3}
\end{figure}

In Figs.\ \ref{fig1}-\ref{fig3} momentum and angular distributions in
the kinematic conditions of the NE18  experiment~\cite{NE18-1} for
carbon are shown for events generated without FSI, i.e. in PWIA. In
all figures the solid line refers to $Q^2=1.04$ (GeV/$c$)$^2$,  
$E_e=2.015$ GeV and the dashed line to $Q^2=6.77$ (GeV/$c$)$^2$,  
$E_e=5.12$ GeV, respectively. 

In Fig.\ \ref{fig1}a at $Q^2=1.04$ (GeV/$c$)$^2$ the momentum
distribution  $N(p_e)$ of electrons is strongly peaked around 1.5
(GeV/$c$)$^2$, while at $Q^2=6.77$ (GeV/$c$)$^2$ it is much broader and
extends over a range of roughly 2 GeV/$c$. The main reason for this
broadening is that at higher electron energies the spreading of the c.m.
energy due to Fermi motion is  much more pronounced. In Fig.\
\ref{fig1}b the momentum distributions $N(p')$ of final protons are
presented. Their shapes are qualitatively similar to those for
electrons, They have, however, different positions of the maxima.

The angular distributions of final electrons ($N(\theta_e)$) and protons
($N(\theta_p)$) are shown in Figs.\ \ref{fig2}a and \ref{fig2}b,
respectively. The angle $\theta_e$ is the scattering angle, while 
$\theta_p$ is defined in the electron scattering plane with respect to a
$\hat z$ axis directed along the incident beam. The INC model reproduces
the expected spread of the angular distributions due to Fermi motion.

The angular distribution $N(\phi_{ep})$ is shown in Fig.\ \ref{fig3} for
events corresponding to protons ejected out of the scattering plane.
In fact, $\phi_{ep}$ is the angle between the electron scattering plane
and the plane defined by the momenta ${\bold p}'$  and ${\bold p}_e$ of
the emitted proton and the incident electron, respectively. If Fermi
motion would be absent, this distribution in the c.m. of the final
system would be described by the delta function $\delta(\phi_{ep}- \pi)$. 
On the contrary, the width of this distribution is determined by the ratio
between the transverse component of the Fermi momentum with respect to
the scattering plane and ${\bold p}'$. At larger $Q^2$, $p'$ is larger
and the distribution becomes narrower. 

\subsection{The effect of FSI}
\label{sec:fsi}

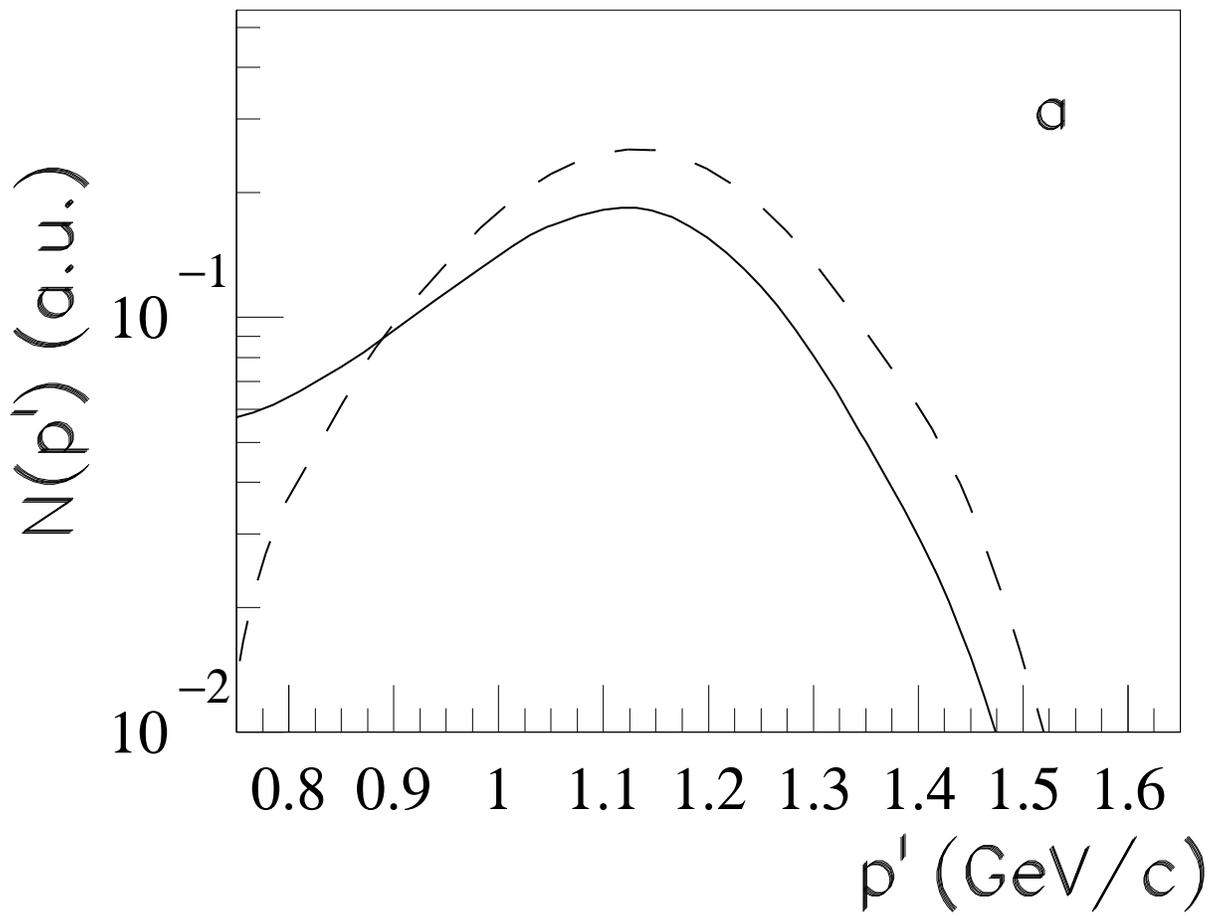
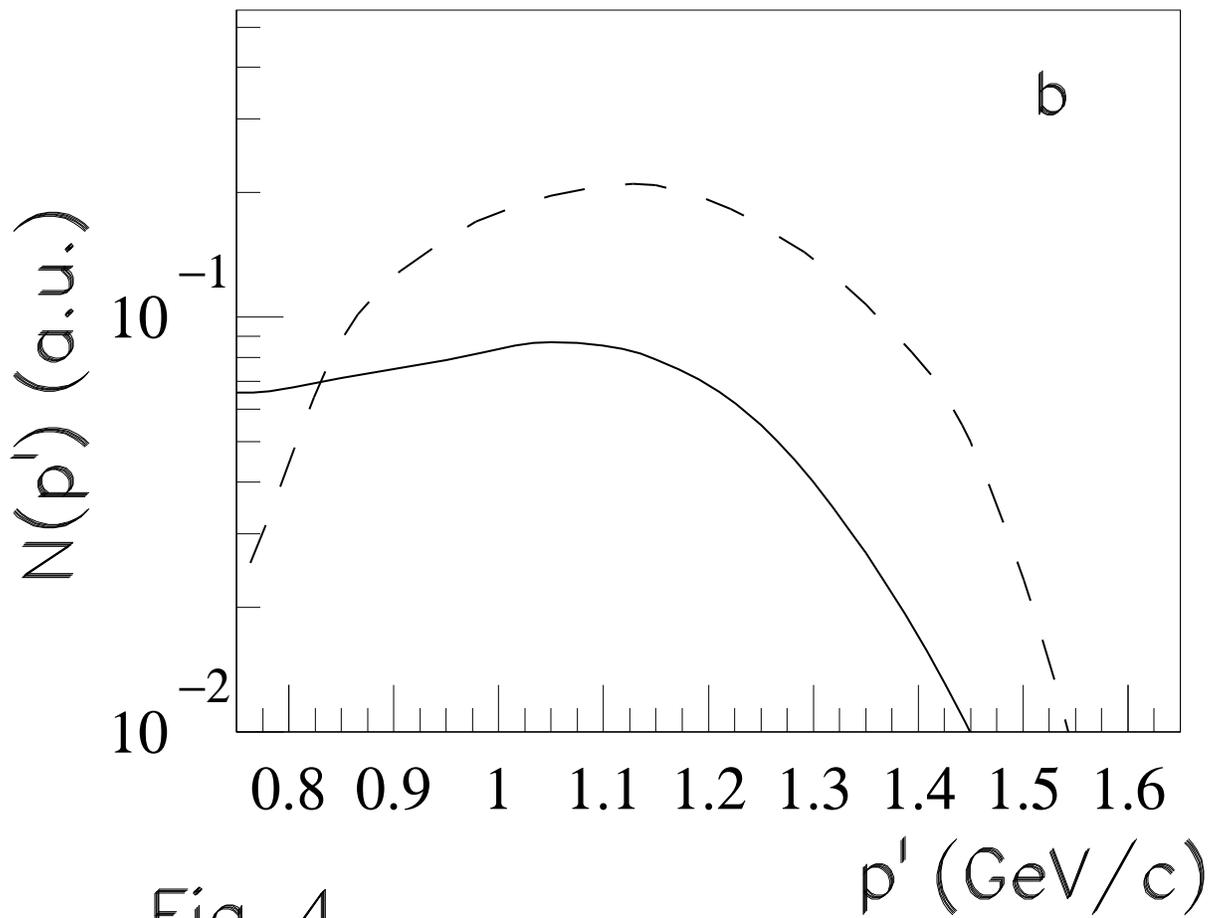
\begin{figure}
\caption{Proton momentum distributions, integrated over the angles 
$\theta_p$ and $\phi_{ep}$ (see text), for the $(e,e'p)$ reaction on C 
(a) and Au (b) at $Q^2=1.04$ (GeV/$c$)$^2$, $E_e=2.015$ GeV. 
The solid and dashed lines are results of the INC model with and without
FSI, respectively.}
\label{fig4}
\end{figure}

Fig.\ \ref{fig4} shows the proton spectrum $N(p')$ integrated over the
angles $\theta_p$ and $\phi_{ep}$ at $Q^2=1.04$ (GeV/$c$)$^2$, $E_e=2.015$ 
GeV in PWIA (dashed line) and with FSI
computed within the INC model (solid line). In the upper part (Fig.\
\ref{fig4}a) the protons are emitted from carbon; therefore, the dashed
line corresponds to the solid line in Fig.\ \ref{fig1}b on a smaller
scale. As expected, the struck proton looses part of its momentum
because of rescattering and pion production. Consequently, FSI make the
spectrum softer and move part of the strength to lower momenta. The
effect is even more pronounced for gold (Fig.\ \ref{fig4}b).

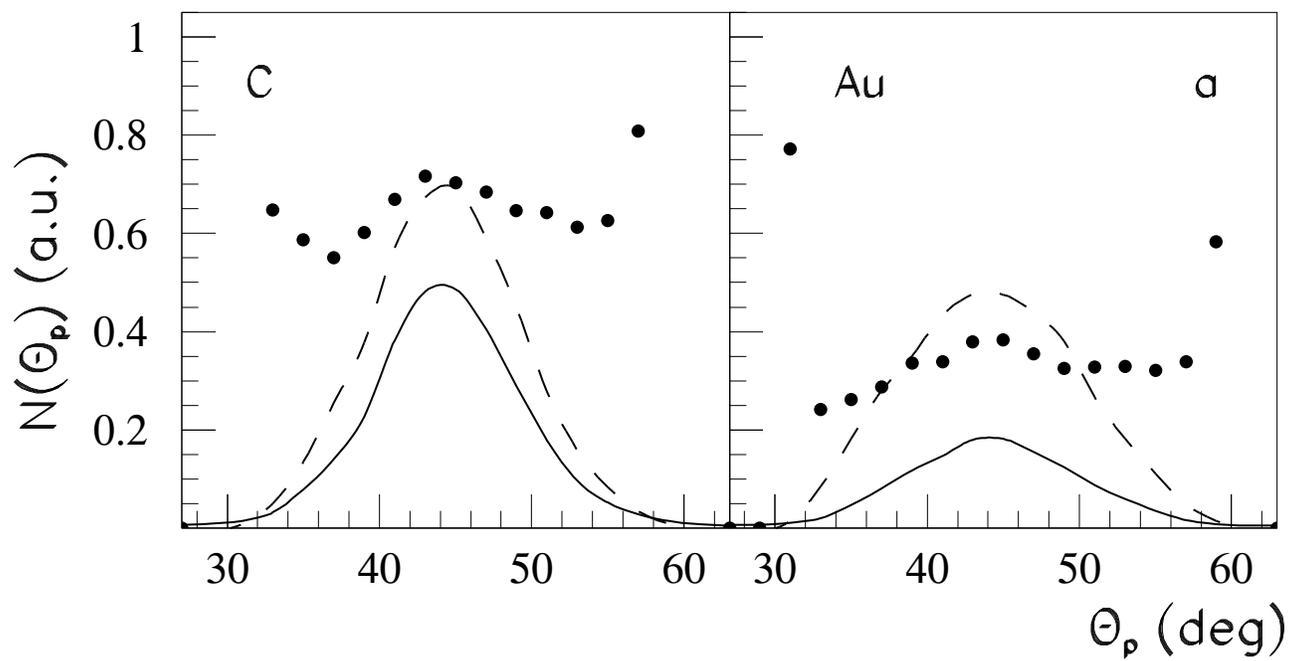
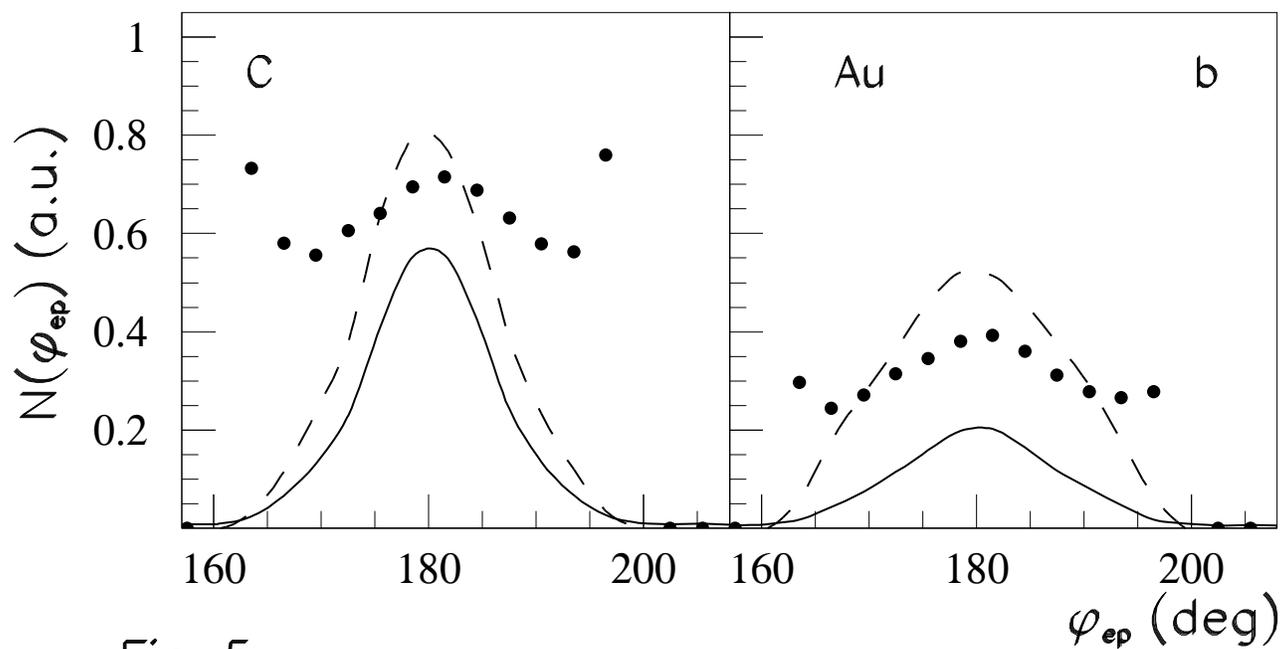
\begin{figure}
\caption{Angular distributions, integrated over the proton momentum
interval $1.1-1.3$ GeV/$c$, for protons detected in the scattering plane 
(a) and out of plane (b) in the $(e,e'p)$ reaction on C (left) and Au 
(right) at $Q^2=1.04$ (GeV/$c$)$^2$, $E_e=2.015$ GeV, 
with the same notations as in Fig.\ \protect{\ref{fig4}}. The solid
circles are the ratio between the results given by the solid
and dashed lines, that is equivalent to the nuclear transparency
coefficient (see text).}
\label{fig5}
\end{figure}

In a very similar manner, the same effect is evident also for the angular 
distributions of final protons scattered in plane
($N(\theta_p)$) and out of plane ($N(\phi_{ep})$), as it is shown in
Fig.\ \ref{fig5}a  and Fig.\ \ref{fig5}b, respectively, for carbon and
gold targets in the same conditions and with the same notations as in 
Fig.\ \ref{fig4}, except that the distribution is integrated over the
interval $1.1 \leq p' \leq 1.3$ GeV/$c$. In this case, FSI redistribute 
the events over a wider angular range because of rescattering. 

Experimental setups usually require kinematic cuts on
momentum, angular and missing momentum/energy distributions. Therefore,
it is important to compare those distributions in cases where FSI are
switched off and on for the same cuts. This can be done in the INC model 
in a natural way. 

The solid circles in Fig.\ \ref{fig5} describe the ratio 
$T_{\text{INC}} = N_{\text{FSI}} / N_{\text{PWIA}}$ between the 
distributions of events with and without FSI, that is actually equivalent 
to NT. In the range of angles around the maximum, the ratio 
$T_{\text{INC}}$, integrated over the other variables, is approximately 
constant. For example, integrating over $\phi_{ep}$ for carbon 
$T_{\text{INC}} \sim 0.6  \div 0.7$
for  $34^{\circ} \leq \theta_p \leq 54^{\circ}$ and for gold 
$T_{\text{INC}} \sim 0.25 \div 0.35$ 
for $34^{\circ} \leq \theta_p \leq 56^{\circ}$. 
Similarly, integrating over $\theta_p$
$T_{\text{INC}}$ gets the same values for carbon and gold, respectively,
in the range $165^{\circ} \leq \phi_{ep} \leq 195^{\circ}$. 

This stability of $T_{\text{INC}}$ over rather wide angular intervals
suggests that it depends mainly on the nuclear density along the
propagation trajectory of the struck proton. If the angular cuts would
be performed inside the above indicated intervals, the size of 
$T_{\text{INC}}$ would be almost independent of the specific choice of
the cuts. As the angular distributions $N_{\text{INC}}$ are broader than
$N_{\text{PWIA}}$, the values of $T_{\text{INC}}$ increase in the tail
regions and have a large uncertainty at those angles, where the proton
yield, calculated in PWIA, is very small and the dominant contribution
comes from rescattering.

The version of the INC model here adopted cannot give completely
realistic distributions in missing momentum (${\bold p}_m = {\bold p}'
- {\bold q}$) and missing energy ($E_m$), because it uses a spectral
function corresponding to the Fermi gas model. However it is
instructive to analyse the $p_m$ dependence of the ratio
$T_{\text{INC}}$ in comparison with the one of the conventional Glauber
approach.

In Ref.~\cite{nikoetal95} it is argued that, after integrating over the 
missing momentum $p_{m_{\scriptscriptstyle \text{T}}}$ transverse 
to the propagation axis, only the inelastic proton-nucleon cross 
section should contribute to the Glauber multiple-scattering 
series, which describes the attenuation of the ejected proton 
flux. The argument is that the elastic cross section leads just to 
a broadening of the $p_{m_{\scriptscriptstyle \text{T}}}$ 
distribution while inelastic rescatterings suppress the ejectile flux at
any $p_{m_{\scriptscriptstyle \text{T}}}$, according to a mechanism
similar to the Gribov's inelastic shadowing~\cite{nikozak75}. 
Since at $p_{m_{\scriptscriptstyle \text{T}}} = 0$ the total 
proton-nucleon cross section contributes, in this framework NT is 
expected to be an increasing function with 
$p_{m_{\scriptscriptstyle \text{T}}}$. 

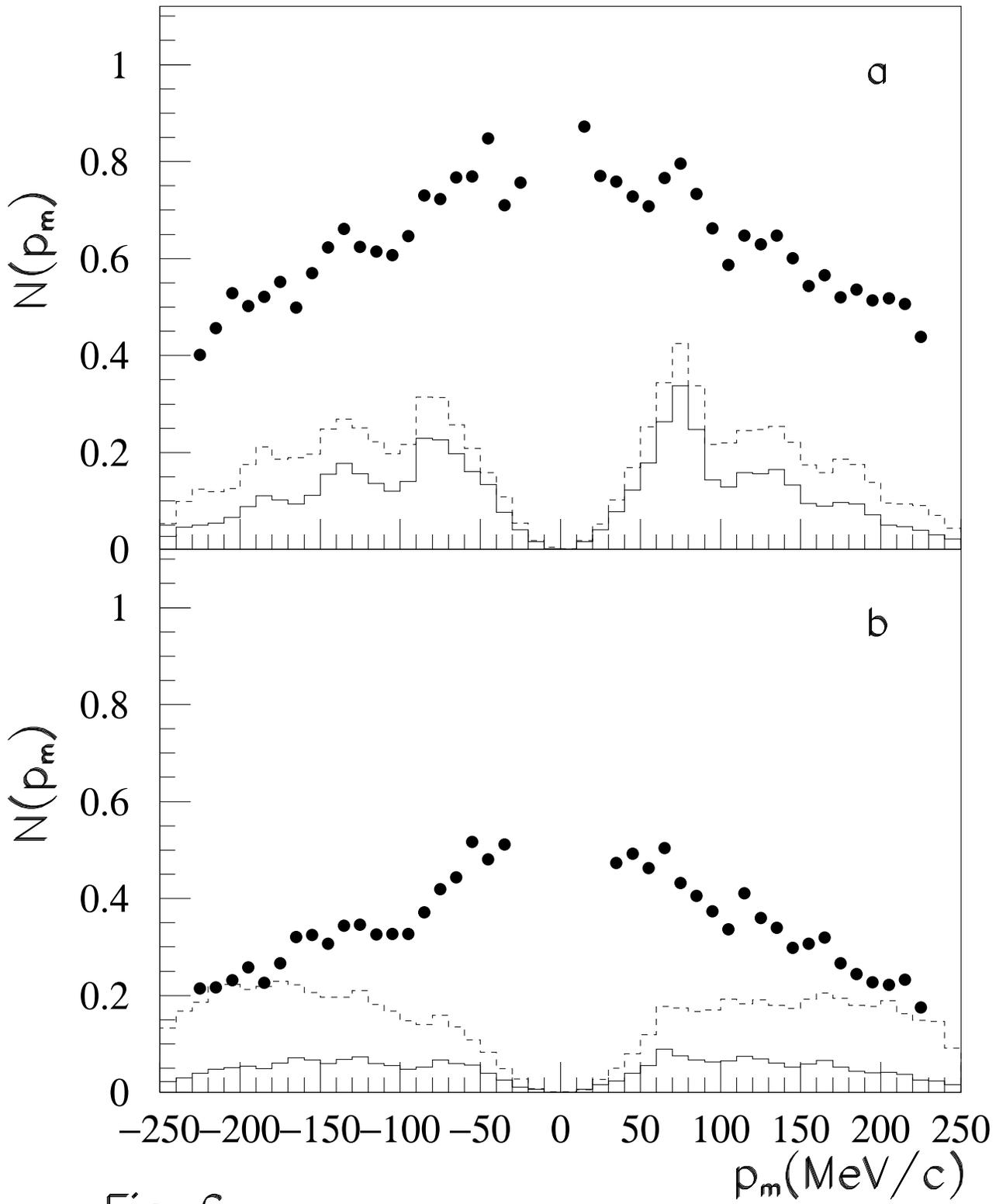
\begin{figure}
\caption{Proton missing momentum distributions and nuclear transparency
coefficient generated by the INC model in the  $(e,e'p)$ reaction on C
(a) and Au (b) at $Q^2=1.04$ (GeV/$c$)$^2$, $E_e=2.015$ GeV, with the same 
notations as in Fig.\ \protect{\ref{fig4}}.}
\label{fig6}
\end{figure}

In Fig.\ \ref{fig6}a and Fig.\ \ref{fig6}b the missing momentum
distributions for carbon and gold targets are shown, respectively, at the
same $Q^2, E_e$ and with the same notations as in Fig.\ \ref{fig4}.
The sign of $p_m$, according to Ref.~\cite{cebaf91013exp}, is defined
positive (negative) when the angle of ${\bold p}'$ with respect to the
incident beam is larger (smaller) than the angle of ${\bold q}$. The
general trend is that, at least at relatively small outgoing proton 
angles, $T_{\text{INC}}$ decreases with increasing $\vert p_m \vert$ 
(and, therefore, $\vert p_{m_{\scriptscriptstyle \text{T}}} \vert$), 
contrary to the previous expectations and in agreement
with Ref.~\cite{br96a} (see, in particular, Fig. 4 therein at angles
corresponding to $p_m$ below the Fermi momentum). A possible 
explanation (confirmed and justified also in the framework of the EA, see 
Sec.\ \ref{sec:output}) relies on the observation that struck protons 
with higher missing momenta mainly come from deeper zones inside the 
nucleus. Therefore they must propagate through larger distances inside 
the nuclear medium before escaping towards the detector. 

Imposing a constraint on the range of explored missing momenta, as in
the NE18 experiment~\cite{NE18-2} where $0 \leq p_m \leq 250$ MeV/$c$, 
could affect the previous argument based on the interference between
elastic and inelastic channels. A quantitative estimate is possible in
the INC model, where all particles can be tagged and recognized at each
step during their propagation inside the nuclear medium. Therefore, one
can compute the number $N_{\text{dir}}$ of events obtained according to
the attenuation of the proton flux in the forward direction and the
number $N_{\text{resc}}$ of events where the protons fell into the
detector acceptance coming from very different initial conditions
due to elastic and inelastic rescatterings. Two conditions were selected
that correspond to $Q^2 = 1.04$ and $6.77$ (GeV/$c$)$^2$ in the NE18
experiment~\cite{NE18-2}, but no cuts were applied on $p_m$ (see Table\
\ref{tabI}). The kinematic restrictions for  $p'$ and $\phi_{ep}$ are
slightly softer than in the NE18  experiment~\cite{NE18-2}, but further
checks at points with higher statistics have shown that the results are
stable againts stronger cuts. As indicated in Table\ \ref{tabI}, the
ratio $R = N_{\text{resc}} / N_{\text{dir}}$ is always small. Therefore,
under the conditions of the NE18 experiment~\cite{NE18-2} the fraction of
indirect protons reaching the detector with large
$p_{m_{\scriptscriptstyle \text{T}}}$ after elastic or inelastic
rescattering is small.

\section{Nuclear transparency in EA}
\label{sec:eik}

In exclusive $(e,e'p)$ reactions on nuclei, where the residual system is
left in a well defined final state,  the basic ingredient of the
calculation is the scattering  amplitude~\cite{bgprep}
\begin{equation}
J^{\mu}_{\alpha} (Q^2, {\bold q}, E_{R_{\alpha}}) = 
\displaystyle{\int} \text{d} {\bold r} \text{d} \sigma 
\text{e}^{{\scriptstyle \text{i}} {\bdsmall {\scriptstyle q}} \cdot 
{\bdsmall {\scriptstyle r}}} 
\chi^{\left( -\right)\, *}_{\bdsmall {\scriptstyle p'}} ({\bold r}, 
\sigma) \, {\hat J}^{\mu} (Q^2, {\bold q}, {\bold r}, \sigma) \  
\phi^{}_{\alpha, E_{R_{\alpha}}} ({\bold r}, \sigma) ,
\label{eq:scattampl}
\end{equation}
where ${\hat J}^{\mu}$ is the nuclear charge-current density 
operator. The scattering wave function 
$\chi^{\left( -\right)}_{\bdsmall {\scriptstyle p'}}$ is the solution 
of a Schr\"odinger equation involving an optical potential $V$ which
effectively describes the interaction between the  residual nucleus,
recoiling with momentum  $- {\bold p}^{}_m$ and mass $M^{}_R$, and the
outgoing proton, detected  in the direction defined by $\cos \gamma =
{\bold p}' \cdot  {\bold q} / p' q$. The proton bound state 
$\phi^{}_{\alpha, E_{R_{\alpha}}}$ is the solution of an eigenvalue problem
involving a single-particle local potential of the Woods-Saxon type,
which also depends on the excitation energy $E^{}_{R_{\alpha}}$ of the 
residual nucleus corresponding to the proton removal from the shell with
quantum numbers $\alpha$. Since  the kinetic energy of the residual
nucleus is given by~\cite{frumou,bgprbook}
\begin{equation}
K^{}_{R_{\alpha}} = \left( p^2_m + 
\left( M^{}_R + E^{}_{R_{\alpha}} \right)^2 
\right)^{1/2} -
M^{}_R - E^{}_{R_{\alpha}}  ,
\label{eq:reskin}
\end{equation}
also the missing energy of the reaction explicitly depends on the
produced hole through the relation
\begin{equation}
E^{}_{m_{\alpha}} = \omega - K^{}_{p'} - 
K^{}_{R_{\alpha}}  .
\label{eq:emiss}
\end{equation}
Therefore, in the following, the complete dependence of the 
scattering amplitude on the bound-state quantum numbers $\alpha$ 
is exploited by the notation 
$J^{\mu}_{\alpha} (Q^2, {\bold p}^{}_m, 
E^{}_{m_{\alpha}})$.

Here our interest is on the properties of the scattering wave
$\chi^{\left( -\right)}_{\bdsmall {\scriptstyle p'}}$ and the simplified
picture is considered retaining just the longitudinal component ${\hat
J}^0$ to leading order $o(1)$ of the nonrelativistic expansion. 
Consequently, the cross section becomes proportional 
to~\cite{br95,br96a,br96b}
\begin{equation}
\Bigg \vert \displaystyle{\int}  \text{d} {\bold r} 
\text{d} \sigma \  \text{e}^{{\scriptstyle \text{i}} {\bdsmall 
{\scriptstyle q}} \cdot {\bdsmall {\scriptstyle r}}} 
\chi^{\left( - \right) \, *}_{\bdsmall {\scriptstyle p'}} ({\bold r}, 
\sigma) \phi^{}_{\alpha, E_{R_{\alpha}}} ({\bold r}, \sigma) 
\Bigg \vert^2 
\equiv S^D_{\alpha} (Q^2, {\bold p}^{}_m, E^{}_{m_{\alpha}})  ,
\label{eq:specdist}
\end{equation}
which is traditionally identified as the ``distorted'' spectral density
$S^D_{\alpha}$~\cite{bgpf79} at the  missing energy $E^{}_{m_{\alpha}}$
of the residual  nucleus with a hole with quantum numbers $\alpha$.

The Schr\"odinger equation for the scattering state can be solved for
each partial wave of  $\chi^{\left( - \right)}_{\bdsmall {\scriptstyle
p'}}$ up to a maximum angular momentum $L_{\text{max}} (p')$, which
satisfies a convergence criterion. The boundary condition is such that
each incoming partial wave coincides asymptotically with the 
corresponding component of the plane wave associated to $p'$. 
Typically, this method has been successfully applied to $(e,e'p)$
scattering with proton momenta below 0.5 GeV/$c$  and $L_{\text{max}} <
50$ for a large variety of complex optical potentials, including also
spin degrees of freedom~\cite{bgprbook}.

At higher energies the Glauber method~\cite{glau59} suggests an
alternative way (based on the EA) to solve the Schr\"odinger equation
by reducing it to a first-order differential equation along the
propagation axis  $\hat z$:
\begin{equation}
\left( {\partial \over{\partial z}} - \text{i} p' \right) 
\chi = {1 \over {2 \text{i} p'}} V \chi  .
\label{eq:schroglau}
\end{equation}
The standard boundary condition requires that asymptotically $\chi
\rightarrow 1$ corresponding to an  incoming unitary flux of plane
waves. By substituting the  solution of Eq.\ (\ref{eq:schroglau}) into
Eq.\  (\ref{eq:specdist}) one gets the final expression for the 
distorted spectral density~\cite{br96b}:
\begin{equation}
S^D_{\alpha} (Q^2, {\bold p}^{}_m, E^{}_{m_{\alpha}}) = \Bigg \vert 
\displaystyle{\int} \text{d} {\bold r} \phi^{}_{\alpha, E_{R_{\alpha}}} 
({\bold r}, \sigma) \times 
\exp \left( - \text{i} {\bold p}_m \cdot {\bold r} \  + \  
\int_{\scriptstyle z}^{\scriptstyle + \infty} V \left( 
{\bold r}_{\scriptstyle \perp}, z' \right) \text{d} z' \right) 
\Bigg \vert^2 . \label{eq:disteik}
\end{equation}
In the pure Glauber model $V(r)$ is determined in a parameter-free way
starting from the elementary free proton-nucleon scattering amplitudes
at the considered energy, while at lower energies, for $(e,e'p)$ 
reactions under quasi-elastic conditions, it usually has a Woods-Saxon
form whose parameters are fixed by fitting the phase-shifts and the
analysing power of elastic (inelastic)  $(p,p)$ scattering on the
corresponding residual nucleus~\cite{bgprbook}.

The EA, whose reliability is supposed to increase with increasing
ejectile energy~\cite{glau59}, has been successfully 
tested~\cite{br95,br96a,br96b} in the momentum range of interest here
($1 \leq p' \leq 6$ GeV/$c$) against the solution of the Schr\"odinger
equation up to $L_{\text{max}} = 120$, as required by the mentioned
convergence criterion. We adopt here the same simple Woods-Saxon form
for the potential $V(r)$, i.e.
\begin{equation}
V(r) = \left( U + \text{i} W \right) \, {\displaystyle {1
\over {1 + \text{e}^{\left( r - R \right) / a}}}} 
\equiv \left( U + \text{i} W \right) \, \rho (r)   ,
\label{eq:opt}
\end{equation}
where $\rho (r)$ is normalized such that $\rho (0) = 1$, $a$ 
is the nuclear diffuseness and $R = 1.2 \times A^{1/3}$ fm. 

At the considered proton momenta, the elementary proton-nucleon
scattering amplitude is dominated by inelastic processes and $V(r)$ is
supposed to be mostly sensitive to the imaginary well depth
$W$~\cite{lech}. However, no  phenomenological phase-shift analysis is
available beyond the inelastic threshold, which could constrain $U$ and
$W$. It has been shown elsewhere~\cite{br96a,br96b} that $S^D_{\alpha}$ 
is rather clearly insensitive to the sign and magnitude of
$U$ for different test values of $(U,W)$, which justifies the choice
$U = 0$, also here adopted. This choice does not contradict the Glauber
model, where the ratio $U/W$ should equal the ratio between the real and
the imaginary parts of the average proton-nucleon forward-scattering 
amplitude, because this ratio is expected to be small anyway above the
inelastic threshold~\cite{lech}.

As suggested by Eq.\ (\ref{eq:schroglau}), the Glauber approach predicts
$W \propto p'$ as far as the proton-nucleon total cross section (and,
consequently, the damping of the proton flux) can be considered constant
for different choices of ${\bold p}' \simeq {\bold q}$, i.e. for small
angles  $\gamma$. However, in order to reproduce the NE18
data~\cite{NE18-2}, a smaller proportionality factor $W/p'$ seems to be
required  with respect to the one indicated by the Glauber 
model~\cite{vijpie92,nikoetal94,jap,fraetal95}. Here, we adopt the 
choice $W = 50 \  p'/1400$ MeV which reproduces the damping, observed in
the NE18 experiment for $^{12}$C at $p' \simeq q =  1.4$
GeV/$c$~\cite{NE18-1}. This choice is equivalent to retaining the full
Glauber method, but assuming a smaller proton-nucleon cross  section in
nuclear matter than in free space~\cite{br96b}.

In order to compare the SLAC data with a theoretical prediction based on
the $S^D_{\alpha} (Q^2, {\bold p}^{}_m, E^{}_{m_{\alpha}})$ of Eq.\ 
(\ref{eq:disteik}), which explicitly depends on the quantum numbers
$\alpha$ of the produced hole and, therefore, refers to a completely
exclusive process, it is necessary to define a theoretical NT
coefficient as follows:
\begin{equation}
T_{\text{EA}} (Q^2) = \displaystyle{
{ {{\lower5pt\hbox{$_{\alpha}$}} \kern-7pt 
   {\hbox{\raise2.5pt\hbox{$\sum$}}} 
   {\lower5pt\hbox{$_{p_m}$}} \kern-10pt
   {\hbox{\raise2.5pt \hbox{$\sum$}}} \  
   S^D_{\alpha} (Q^2, {\bold p}^{}_m, 
    E^{}_{m_{\alpha}})} 
  \over
  {{\lower5pt\hbox{$_{\alpha}$}} \kern-7pt 
   {\hbox{\raise2.5pt\hbox{$\sum$}}} 
   {\lower5pt\hbox{$_{p_m}$}} \kern-10pt
   {\hbox{\raise2.5pt \hbox{$\sum$}}} \  
   S^{PW}_{\alpha} (Q^2, {\bold p}^{}_m, 
    E^{}_{m_{\alpha}})}  }  } .
\label{eq:tras}
\end{equation}
Eq.\ (\ref{eq:tras}) gives the ratio between the nuclear responses 
$S^D_{\alpha}$ and $S^{PW}_{\alpha}$ obtained with and without FSI,
respectively, for each $Q^2$ incoherently summed over the range of
proton angles $\gamma$ covered by the NE18 experiment (corresponding to
different  $p^{}_m$~\cite{NE18-2}) and over the quantum numbers $\alpha$ 
of the occupied shells in the considered target nucleus.

\section{Comparison with data}
\label{sec:output}

Experimental data for NT in quasielastic $A(e,e'p)$ reactions are
available from BATES~\cite{bates}, SLAC~\cite{NE18-1,NE18-2} and, in a
preliminar form, from TJNAF~\cite{cebaf91013exp}. They are obtained by
taking the ratio between the sum over the observed events in the selected
kinematic region and the corresponding theoretical quantity calculated
in PWIA for the same region, except for the BATES experiment where the
ratio between exclusive and inclusive cross sections was taken. The
data cover the range  $0.3 \leq Q^2 \lesssim 7$ (GeV/$c$)$^2$. 

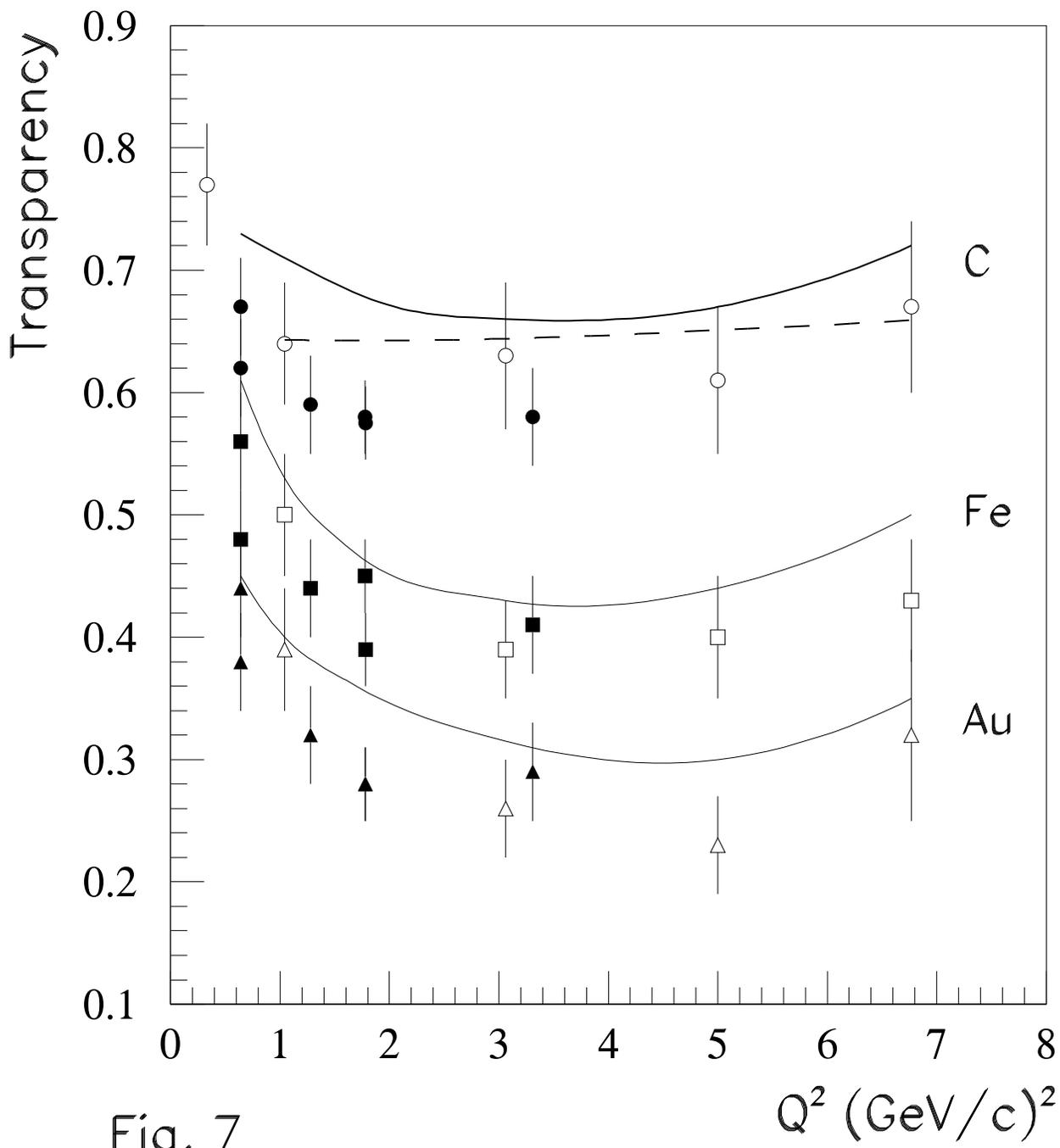
\begin{figure}
\caption{Nuclear transparency, integrated over missing momentum and
energy, as a function of $Q^2$ for the $A(e,e'p)$ reaction. Open
symbols are data from the NE18 experiment at
SLAC,
 but for the point
at $Q^2 = 0.3$ (GeV/$c$)$^2$ obtained at BATES.
Solid symbols are the preliminar data taken at
TJNAF.
Circles, squares and triangles refer
to carbon, iron and gold targets, respectively. The solid lines are
results of the INC model, while the dashed line is obtained in the EA for
carbon.}
\label{fig7}
\end{figure}

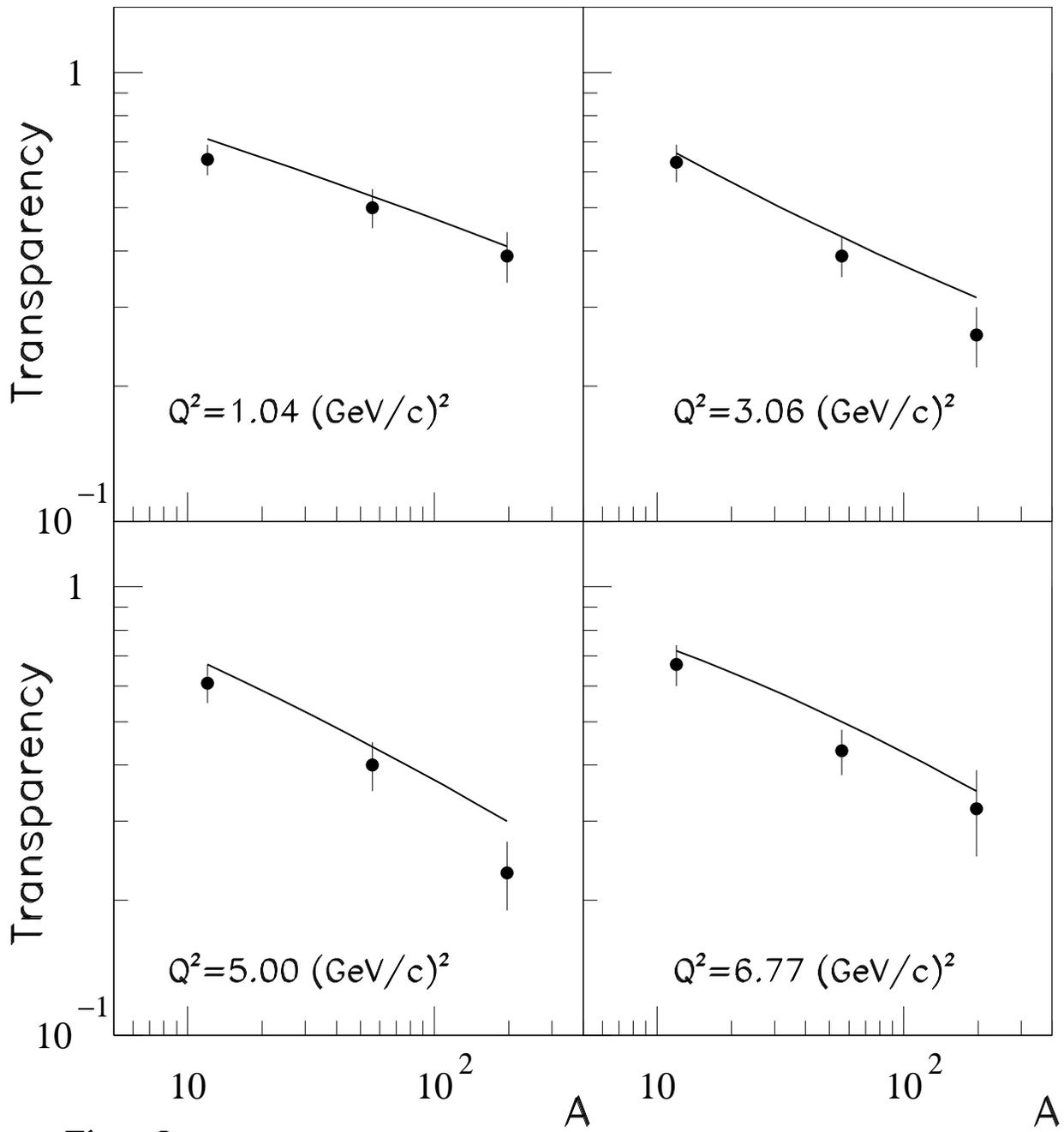
\begin{figure}
\caption{Nuclear transparency, integrated over missing momentum and
energy, as a function of $A$ and $Q^2$. Data are from the NE18 experiment 
at SLAC. 
The solid lines are the result from the INC model.}
\label{fig8}
\end{figure}

In Fig.\ \ref{fig7} open symbols refer to the NE18 experiment performed
at SLAC, with the  exception of the point at $Q^2 \sim 0.3$
(GeV/$c$)$^2$ that has  been obtained at BATES. Solid symbols indicate
the preliminar data from TJNAF. From top to bottom, circles, squares
and  triangles give the results for carbon, iron and gold targets, 
respectively. Theoretical calculations of $T_{\text{INC}}$ in the
framework of the INC model are indicated by solid lines. They implement
all the experimental cuts in angles and momenta as well as the
integration over missing momenta and energies covered by the NE18
experiment~\cite{NE18-1,NE18-2}. Agreement with data is quite
satisfactory and is confirmed in Fig.\ \ref{fig8}, where the
$A$ dependence of $T_{\text{INC}}$, integrated over missing
momentum and energy, is shown for fixed values of $Q^2$. 

For sake of comparison, in Fig.\ \ref{fig7} the dashed line shows the
result of $T_{\text{EA}}$ obtained for carbon after summing over its
occupied $s\textstyle{1 \over 2}$ and $p\textstyle{3 \over 2}$ shells in 
Eq.\ (\ref{eq:tras}) as well as
over $p_m$ in the range corresponding to the proton angles measured in
the NE18 experiment~\cite{NE18-1}. In fact, in the fixed kinematics of an
exclusive reaction there is a one-to-one correspondence between $p_m$
and $\theta_p$ (or, equivalently, $\gamma$). Agreement with data is very
good. Also the similarity between the results of two completely different
models is remarkable. 

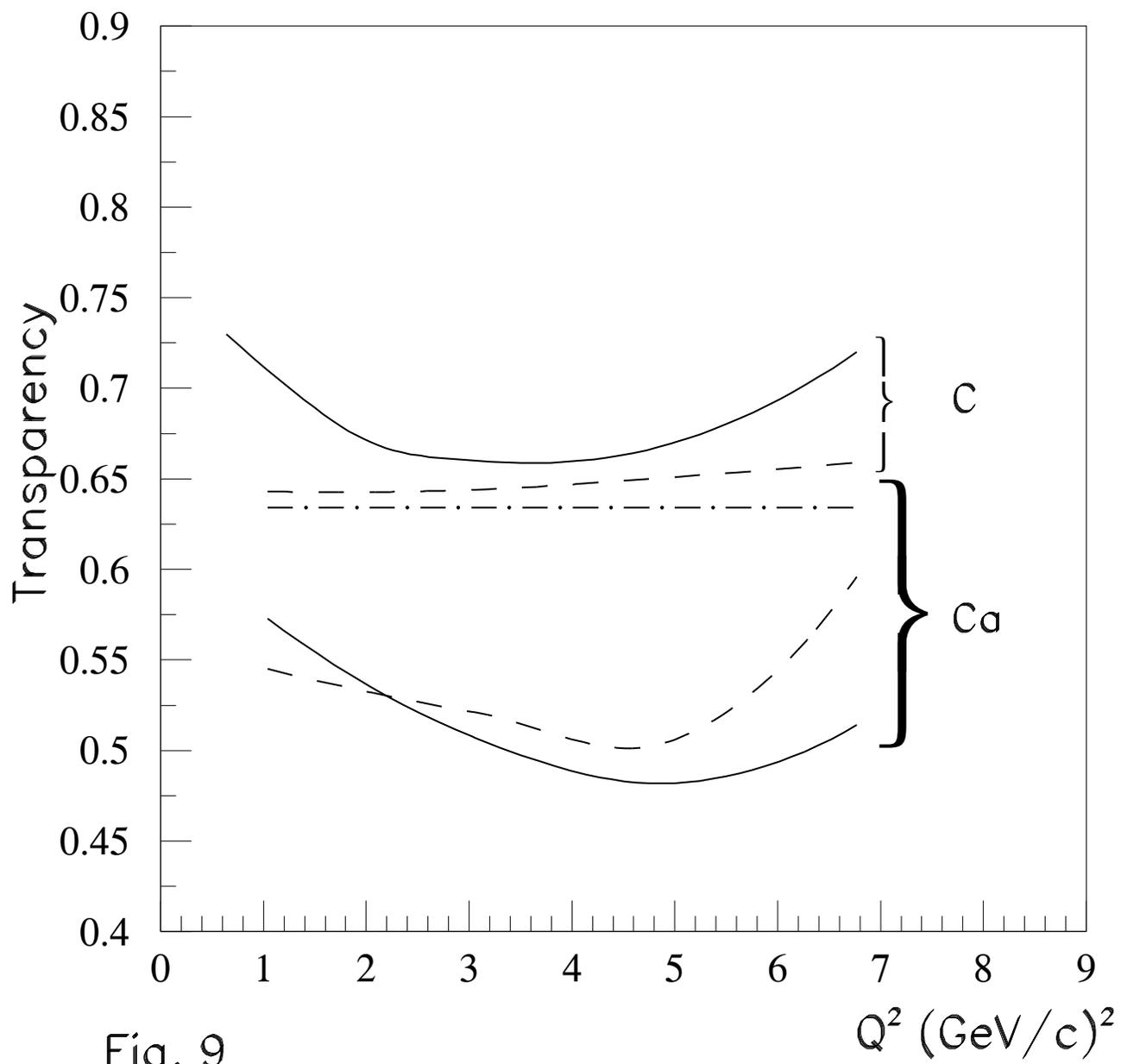
\begin{figure}
\caption{Nuclear transparency, integrated over missing momentum and
energy, as a function of $Q^2$ for the $A(e,e'p)$ reaction on
$^{12}$C and $^{40}$Ca targets. The solid and dashed lines are  
results of the INC model and of the EA, respectively. The dot-dashed line
is the result of the EA for $^{40}$Ca with $\gamma = 0$ at all values
of $Q^2$ (see text).}
\label{fig9}
\end{figure}

This is confirmed in Fig.\ \ref{fig9}, where the comparison between the INC 
model and the EA is extended also to the $^{40}$Ca target. The shape of 
$T_{\text{EA}}$ (indicated by the dashed line) is essentially given by the 
fact that, according to the NE18 experimental setup, for each $Q^2$ different
ranges are covered for the proton angles, and consequently for the
missing momentum $p_m$. The different shells, then, contribute
differently with their $p_m$ dependence so that at each
$Q^2$, according to the selected range of $p_m$, the relative weight of
their contribution is changing. As a test, in Fig.\ \ref{fig9} the 
dot-dashed line is also shown, which refers to NT for the $^{40}$Ca$(e,e'p)$ 
reaction in the same kinematics of the NE18 experiment but keeping the 
outgoing proton angle $\gamma = 0$ at each value of $Q^2$: keeping the same
proton angle makes NT independent of $Q^2$, at least in the observed
range. 

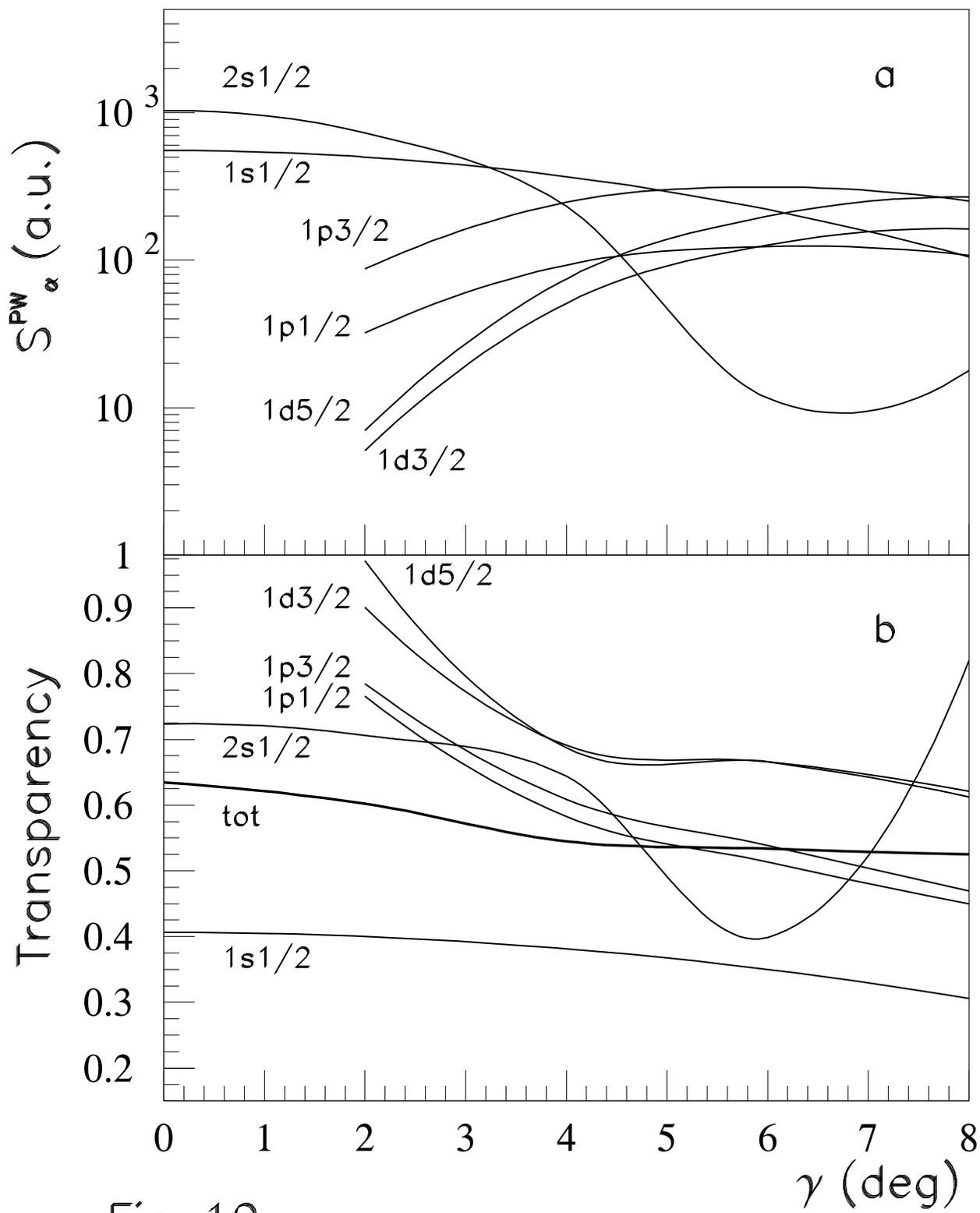
\begin{figure}
\caption{Angular distributions of the PWIA nuclear response (a) and of the 
nuclear transparency calculated in the EA framework (b) of each target 
shell for the $^{40}$Ca$(e,e'p)$ reaction at $p' = q = 1$ GeV/$c$. The 
resulting total nuclear transparency of Eq.\ (\protect{\ref{eq:tras}}) is 
also plotted in the lower part (b) and labelled ``tot''.}
\label{fig10}
\end{figure}

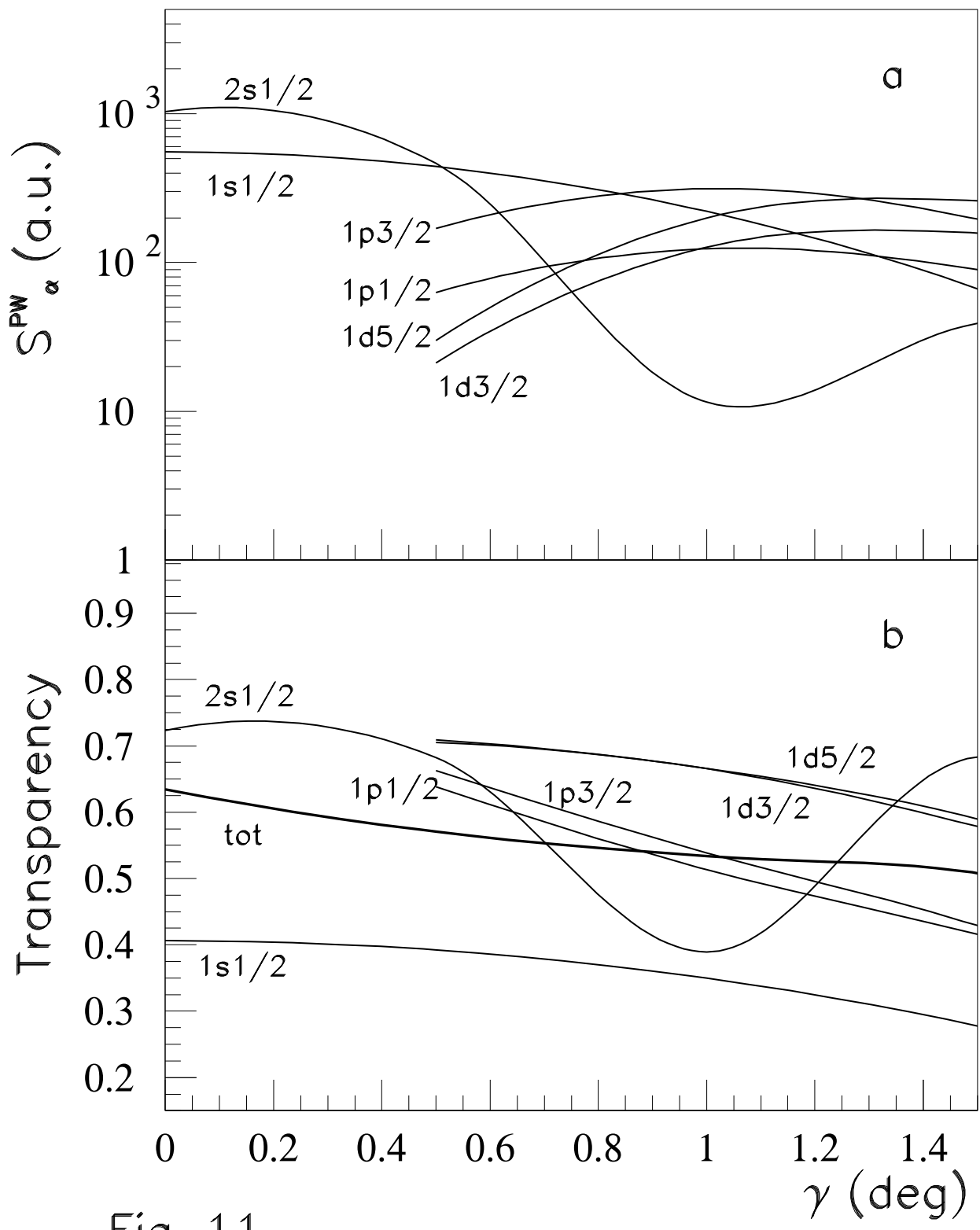
\begin{figure}
\caption{The same as in Fig.\ \protect{\ref{fig10}} but for $p' = q = 6$ 
GeV/$c$.}
\label{fig11}
\end{figure}

The exclusive nature of direct knockout, intrinsic in the definition of 
$T_{\text{EA}}$ in Eq.\ (\ref{eq:tras}), allows for a more detailed
analysis of the contribution of each shell to the integrated transparency as
well as to its angular distribution. In Fig.\ \ref{fig10}a (upper part) 
the PWIA nuclear response $S^{PW}_{\alpha}$, obtained from Eq.\ 
(\ref{eq:disteik}) without FSI, is shown as a function of the proton angle 
$\gamma$ for the $^{40}$Ca$(e,e'p)$ reaction at $p' = q = 1$ GeV/$c$. The 
labels refer to the quantum numbers of the shells building up the structure 
of $^{40}$Ca. At very forward angles protons only come from $s$ shells. At 
higher angles the $2s\textstyle{1 \over 2}$ contribution is irrelevant and 
protons with a non negligible $p_{m_{\scriptscriptstyle \text{T}}}$ come 
from $p$ and $d$ orbitals, as well as from $1s\textstyle{1 \over 2}$. In 
Fig.\ \ref{fig10}b (lower part) the corresponding NT calculated in the EA
framework is shown for each shell as a function of $\gamma$ in the same 
conditions. The curve labelled by ``tot'' refers to the angular 
distribution of the total transparency for the $^{40}$Ca target. The 
angular dependence of the total NT is determined by the dominant 
contribution of the individual shells at each specific angle. Therefore, 
at low angles the total result is due to a delicate interplay between 
$2s\textstyle{1 \over 2}$ and $1s\textstyle{1 \over 2}$ shells, whose 
transparencies are very different. 
At higher 
angles, the total result approximately follows the NT of the $p$ and $d$ 
orbitals. Globally, the total NT is a decreasing function of $\gamma$, or, 
equivalently, of $p_{m_{\scriptscriptstyle \text{T}}}$, in agreement with 
the findings of Sec.\ \ref{sec:fsi} described in Fig.\ \ref{fig6}. The 
same arguments apply to Fig.\ \ref{fig11}, where the angular range explores
the same range of $p_m$ as in Fig.\ \ref{fig10}, but at 
$p' = q = 6$ GeV/$c$. The NT property of being a decreasing 
function of $p_{m_{\scriptscriptstyle \text{T}}}$ is even more evident.
Large variations of NT with the proton emission angle $\gamma$ are then
possible. The larger $Q^2$ and $p'$, the smaller is the $\gamma$ 
corresponding to the same $p_{m_{\scriptscriptstyle \text{T}}}$.
Therefore, within the experimental acceptance $(2^{\circ})$~\cite{NE18-1}
the corresponding angular averaging could miss significant variations of
NT.

In both Figs.\ \ref{fig10} and \ref{fig11} the $p$ and $d$ angular
distributions do not start from $\gamma = 0$, because the PWIA
result is vanishing and, therefore, not contributing to the 
$T_{\text{EA}}$ of Eq.\ (\ref{eq:tras}), while producing an artificial
infinity in the transparency of the single shell at that angle. 

Finally, comparison of Figs.\ \ref{fig10}b and \ref{fig11}b shows 
that integrating over $\gamma$ the curves labelled by ``tot'' (therefore,
integrating them over the same range of $p_m$ reached at different $Q^2$) 
will produce the same total NT coefficient, in agreement with the 
dot-dashed line of Fig.\ \ref{fig9}. Therefore, as previously
anticipated, the $Q^2$ dependence shown by solid and dashed lines in Fig.\
\ref{fig9} and by data in Fig.\ \ref{fig7} can be interpreted, in the
framework of the EA, as a kinematic effect related to the shell structure 
of the target. At different $Q^2$, probing different $p_m$ means probing
different relative weights of each shell contributing to the total NT; 
FSI will be less (more) effective producing an increasing (decreasing)
transparency.

\section{Conclusions}
\label{sec:end}

Nuclear transparency in exclusive quasielastic $A(e,e'p)$ reactions 
has been investigated. Final-state interactions have been treated within 
the intranuclear cascade model and the eikonal approximation. The INC model
describes the available data on NT up to $Q^2\sim 7$ GeV$^2$/$c^2$
rather well without the need of free parameters. Our analysis shows that
the Pauli blocking seems to be the most crucial ingredient and
suppresses the otherwise important interference between coherent and
incoherent rescatterings~\cite{nikoetal94,nikoetal95,nikoetal96}, while 
short-range correlations at such $Q^2$ are less important. The
results of the INC model are also in qualitative agreement with those
obtained in the EA.

In this framework, the $Q^2$ behaviour of NT can be interpreted as a
kinematic effect related to the fact that for each $Q^2$ different ranges 
of missing proton momentum are explored according to the experimental 
setup. In fact, because in the EA the definition of NT is based on a 
genuine exclusive cross section, at each $Q^2$ it is possible to analyse 
the angular distribution not only of NT, but also of the contribution of 
each single target shell. It turns out that NT at small proton angles is 
due to the emission from $s$ shells, while at larger angles shells with
higher angular momentum are important. The different FSI make NT a 
decreasing function of the proton angle (or, equivalently, of the 
transverse missing momentum $p_{m_{\scriptscriptstyle \text{T}}}$) and 
large variations of the NT coefficient are possible 
within the presently available experimental angular acceptance. The 
relative angular contribution of each shell depends on 
$p_{m_{\scriptscriptstyle \text{T}}}$. If the range of explored 
$p_{m_{\scriptscriptstyle \text{T}}}$ is kept constant, the transparency 
coefficient, integrated over the proton angles, does not show any 
$Q^2$ dependence.

In the kinematic conditions presently investigated nuclear transparency 
seems under control. In order to test the onset of other transparency
mechanisms as a function of $Q^2$, our analysis shows that it is
important to keep constant the range of missing momenta covered by the 
experiments.

\acknowledgements

Two of us (Ye. G. and L.A. K.) are grateful to Dipartimento di 
Fisica Nucleare e Teorica of the University of Pavia for kind 
hospitality. This work was supported by INFN (Italy), by the 
International Association for the Promotion of Cooperation with 
Scientists from the Independent States of the Former Soviet Union 
(grant no. INTAS-93-79 Ext.) and performed in part under the
contract ERB FMRX-CT-96-0008 within the frame of the Training and
Mobility of Researchers Programme of the Commission of the European
Union.
              


\begin{table}
\caption{The ratio $R = N_{\protect{\text{resc}}} / 
N_{\protect{\text{dir}}}$ between events for rescattered 
($N_{\protect{\text{resc}}}$) and directly attenuated 
($N_{\protect{\text{dir}}}$) protons for the $A(e,e'p)$ reaction on
C, Fe and Au targets in the kinematics of the NE18
experiment 
but without cuts on $p_m$ (see text).}
\label{tabI}
\begin{tabular}
{c|c|c|c|c|c}
$Q^2$ & $\theta_e$ & $p'$ & $\theta_p$ & $\phi_{ep}$ & $R$ \\
(GeV/$c$)$^2$ & deg & GeV/$c$ & deg & deg & \\
\hline
1.04 & 32-39 & 1.1-1.3 & 40-53 & 170-190 & 2.6 $\%$ (C), 4.2 $\%$
(Fe), 5.7 $\%$ (Au) \\
6.77 & 56.1-57.1 & 4.4-4.6 & 15.5-17.5 & 170-190 & $\leq$ 0.4 $\%$
\\
\end{tabular}
\end{table}


\end{document}